%% file: udq.tex
\documentclass[12pt]{articlo}
\usepackage[dvips]{epsfig}
\usepackage{ific_title}
\begin{document}
\thispagestyle{empty}
\ULname
\docnum{ATLAS \phantom{XXX}\\INDET--NO--xx \phantom{XXX}\\
arch-ive/9606012 \phantom{XXX}}
%hep--ex/96060xx \phantom{XXX}}
\docnum{ June 21, 1996 \phantom{XXX}}
%\docnum{DRAFT--2.X \phantom{XXX}}
%
\vspace*{1cm}
\title{PERFORMANCE OF THE  ATLAS-A SILICON DETECTOR WITH ANALOGUE READ
  OUT}
\author{P.P. Allport, P.S.L.  Booth, C. Green, A.Greenall, M. Hanlon,\\
J.N.  Jackson, T.J.  Jones, J.D.  Richardson,
S.  Mart\'{\i}  i  Garc\'{\i}a\instref{salva},\\
U. Parzefall,  A.E.  Sheridan, N.A. Smith } 
\docinfo{Physics Department, Oliver Lodge laboratory, 
University of Liverpool, \\
Oxford Street, Liverpool L69 3BX, U.K.}
\instfoot{salva}{e-mail address for correspondence: martis@hep.ph.liv.ac.uk}
\abstract{ The performance of an  ATLAS-A silicon micro-strip detector
  prototype with     FELIX   128 analogue  read   out    chip has been
  studied. The  noise level and  the  signal to noise ratio  have been
  measured as a  function of both  detector bias and  temperature.  No
  evidence of micro-discharge  was observed for detector bias voltages
  up to 300 V.}
%
%\comentari{Deadline for submitting comments 21/06/1996}
%\comentari{This is a test version}
%
%\maketitle
%
\parskip    0.2cm
\setcounter{page}{0}
\normalsize
\input{sec_1.dos}

\bibliographystyle{unsrt}
\bibliography{sidet_db}
\end{document}

%% file: sec_1.dos
\section{Introduction} 

The inner tracker   of the ATLAS  experiment  at LHC will  be built of
silicon micro-strip  detectors \cite{atlas_tp}.  These  detectors have
to operate in a high radiation level environment \cite{rad_dose}.  The
effect of the radiation damage is basically an increase of the leakage
current and a  change  in the  effective bulk  doping concentration.  
Therefore,  a reduction of  the signal   to noise  ratio and  detector
efficiency   could  be expected.   In  order  to recover  the expected
performance after  the  detector  has been  irradiated,  the  simplest
solution is to increase the bias voltage.  However, some phenomena can
limit the maximum  value of the  bias voltage applied on  the detector
\cite{v_limit}.   Then,   the degraded  performance of  an  irradiated
detector could stand below the   required minimum for a   satisfactory
operation of the ATLAS tracking system.

In the  case of AC-coupled   silicon detectors, {\sl micro-discharges}
\cite{udq-japones} at the edges of the strips are one of the possible
causes limiting the  maximum value of the  bias voltage applied on the
detector.  The micro-discharges are  manifested as a steep increase in
the noise level with an associated increase in the detector current.

The  ATLAS-A detector  design  is  based  in  $n^+$-type strips on   a
$n$-type bulk,  because   after radiation  damage    the lightly doped
$n$-type bulk inverts to a $p$-type  bulk with an effective $p$ doping
that increases with  dose.  This leads  to high values  of the voltage
needed to fully deplete the detector. This therefore could give a rise
to   micro-discharges.   

In  this note, the possibility    of producing micro-discharges on  an
ATLAS-A silicon  detector  prototype has been  tested.  Therefore, the
prototype performance has been  studied at different biasing  voltages
(from   0  up  to  300 V)  and   at two  different temperatures  ($\pm
20^\circ$C).

The organization of this note is as follows: in section 2 the detector
geometry   and  read out    electronics   are presented.     Then  the
experimental set-up  is described in section 3.   In sections 4 and 5,
the the experimental results in terms of the signal to noise ratio and
the   micro-discharge   production  are    discussed.   Finally,   the
conclusions are summarized in section 6.

\section{Prototype and Read Out Chip electronics \label{sec:electronics}}
The aim of this section is to describe the ATLAS-A silicon micro-strip
detector, its technology  and geometry, and the  basic features of the
FELIX 128 channel analogue read out chip.
\begin{enumerate}
\item The detector used in our studies is a 6 cm $\times$ 6 cm, with a
  56.25 $\mu$m diode pitch (with  an isolation strip) and 112.5 $\mu$m
  read out pitch  (see   figure  \ref{fig:cross}).   The  strips   are
  $n^+$-type  in  a  $n$-type bulk, AC    coupled.   These strips  are
  connected together  to  a bias ring   through a polysilicon resistor
  (600 K$\Omega$).  When biasing  the detector  the $n^+$ strips  were
  directly grounded  and  the $p$-type  backplane was  connected  to a
  negative bias voltage.  Both sides of the detector are surrounded by
  multiple  guard  rings.  The guard  rings  are  used to   shield the
  sensitive part of the detectors and to provide a gradual drop of the
  edge potential, even after irradiation.  Between the $n^+$ strips, a
  $p$-stop  isolation strip is  used to avoid  the short  of the $n^+$
  strips by accumulating positive charge in the SiO$_2$ layer.
  
  From the point of view of a possible production of micro-discharges,
  it is important to know the relative position of the implant and the
  metal  contact  \cite{udq-japones}.  In  this case, the  size of the
  metal contact is smaller by 2$\mu$m (see figure \ref{fig:cross}).
  
\begin{figure}[hbt]
 \centering
 \mbox{\epsfig{file=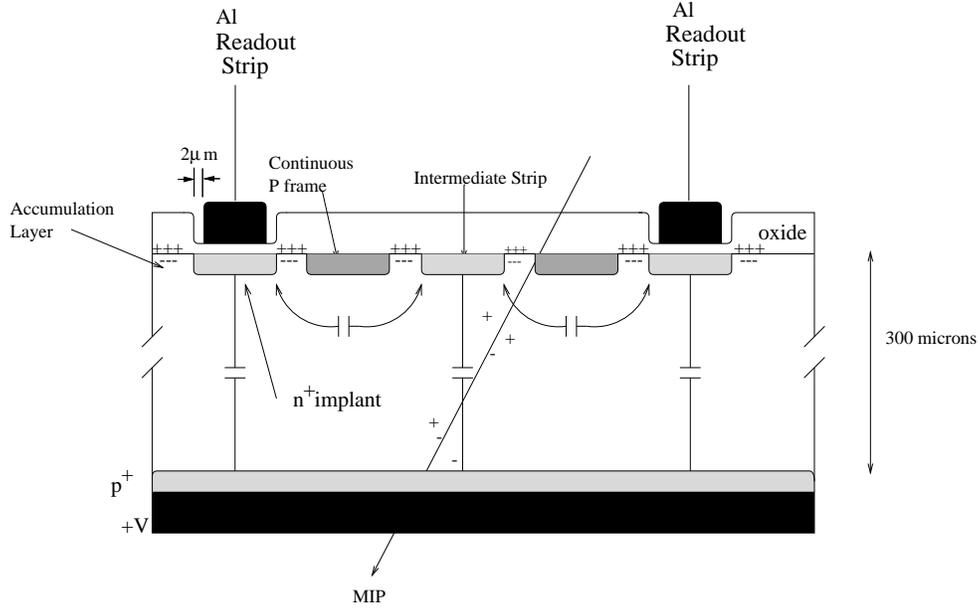,height=8.0cm}}
 \caption{Cross section view of the ATLAS-A prototype.}
 \label{fig:cross}
\end{figure}

\item The FELIX chip \cite{felix}  is a 128 channel integrated circuit
  designed to  read   out silicon  strip detectors  at  the  LHC.  The
  architecture of  the device can be  broken down into  3 main blocks:
  \subitem  $-$  Front  end  pre-amplifier  and  shaper  with a  CR-RC
  characteristic  of  75 ns peaking   time.  \subitem  $-$ An analogue
  pipeline,  1.6$\mu$s deep (Analogue  Delay Buffer).  \subitem $-$ An
  Analogue Pulse Signal Processor (APSP).
  
  The  chip  has been designed   to run in   two modes:
  \subitem {\sl Deconvolution}: the charge  collected is obtained from
  a suitable weight  \cite{deconv} of three  25 ns consecutive  points
  measurements.    It allows to  extract  the  charge deposition in  a
  single event, even in the  case of pile-up  of events. Therefore  it
  can operate at high luminosity running at LHC.
  
  \subitem  $-$ {\sl Peak}: The charge  collected is obtained for each
  event from the measurement  of the height of  the signal  pulse.  It
  requires a minimum distance of 75 ns between two consecutive pulses,
  in order  to  avoid the    pulse pile-up.   Therefore  this mode  is
  inadequate for high luminosity  running where the associated trigger
  rate is 25 ns.

\end{enumerate}

\section{Experimental Set-up}
In this section the experimental set-up used for this study is described.
It consists of the following three main parts:
\begin{itemize}
\item[--] Trigger system,
\item[--] Data Acquisition System (DAS),
\item[--] detector and FELIX biasing.
\end{itemize}
\subsection{The Trigger System}
\begin{figure}[hbt]
 \centering
 \mbox{\epsfig{file=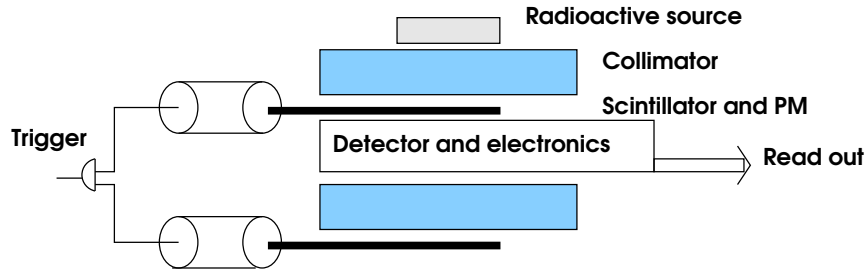,height=3.4cm}}
 \caption{Schematic view of the trigger system}\label{fig:trigger}
\end{figure}

The  trigger system is based on  the coincidences of two scintillators
located  above and below the  silicon micro-strip detector (see figure
\ref{fig:trigger}).  The coincidences can be  triggered either by  the
$\sim 2$ MeV electrons of a radioactive source  or by cosmic rays.  In
the second case a lead shield (used to stop  the soft component of the
cosmic rays \cite{pdg}) replaces the radioactive source.  Then, mostly
single  muons  (minimum   ionizing  particles,  m.i.p.)   trigger  the
scintillators coincidence.

\subsection{Data Acquisition System}

The FELIX read  out uses a VME Test   System which is based  around an
MVME 167  Processor, with OS9 operating system.   Within the VME crate
there is a VME sequencer (SEQSI), a SIROCCO card  with a flash ADC and
a Random Trigger Card.

All  clocks  and  control  signals for   the  FELIX  and  the Analogue
Multiplexor are generated by the sequencer. The Random Trigger Card is
required to   synchronise  external  triggers  (e.g.   the coincidence
trigger from the scintillators) with the  40MHz BCO clock.  Adjustment
of the timing of this synchronised  trigger relative to the data being
collected  is by   means  of a  programmable   delay  unit.  The  data
acquisition is by means of the VME SIROCCO  where the analogue data is
digitized and then written to disc.

\subsection{Detector and FELIX Biasing} 
A Keithley 237 Source-Measure-Unit was used to supply a stable voltage
for the   detector biasing, and  also  allows  the monitoring  of  the
detector leakage  current. Power supplies  for the FELIX were provided
by several low  voltage bench power  supplies, which also provided the
biasing for  the FELIX pre-amplifiers.  It  has been  noticed that the
performance of the FELIX 128 chip is  strongly linked to variations on
its bias  power supplies.  The  main  effects are: an  increase in the
noise level by  a factor of  $\sim 2$, and/or   a global shift  of the
pedestal values of all the channels.

\section{Detector Performance}

In this section  the study of the  detector performance  is presented.
Two types of particles were used to extract the signal to noise ratio.
The  first type is  the muons from  cosmic rays (considered as m.i.p).
However the small trigger rate ($\sim 15 \mu/$hour) of these particles
makes  difficult to   collect a large   amount  of events  in order to
perform  precise  measurements.  Thus,  a  radioactive source emitting
electrons of $\sim 2$ MeV was used instead.

A full range of bias voltages (from 30 V,  just after depletion, up to
300  V)  and two different  temperatures  (room $\sim +20^\circ$C, and
$-20^\circ$C) were  used to determine  the main factors  affecting the
resolution of a silicon micro-strip detector:
\begin{itemize}
\item the noise level,
\item the signal to noise ratio,
\item the cluster size.
\end{itemize}

\subsection{Noise Studies \label{sec:noise}}

\subsubsection{Inter Strip Capacitance \label{sec:is_cap}}
The inter strip capacitance is one of the most important parameters of
the detector performance in terms of the noise level, and consequently
in establishing the signal to noise ratio. In  the cases of AC-coupled
detectors,   with     integrated coupling   capacitors,  the  measured
capacitance is a convolution of  both: the coupling capacitors and the
direct capacitance between the implant layers.

\begin{figure}[hbt]
  \centering \mbox{\epsfig{file=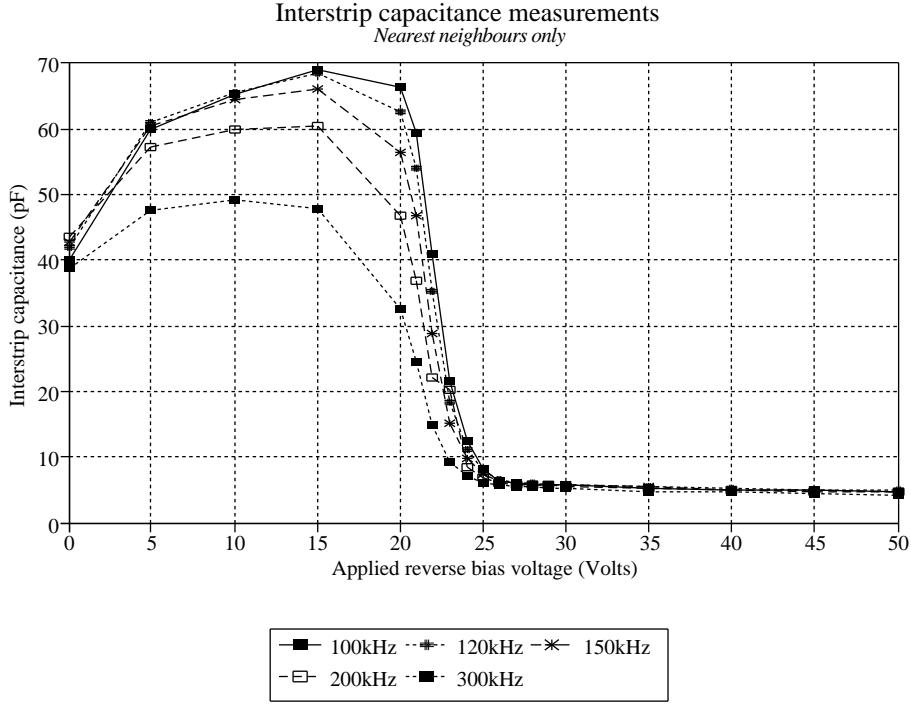,height=9.9cm}}
 \caption{Inter strip capacitance to the nearest neighbours.}
 \label{fig:is_cap}
\end{figure}

The  measured values for the  inter strip capacitance  of one strip to
its two nearest  neighbours, as  a function  of the bias  voltage, are
shown in figure \ref{fig:is_cap}.  It  is clearly  seen how the  inter
strip capacitance is  drastically reduced when  the detector is  fully
depleted ($<$25V).   It  falls to 5 pF,  (equivalent  to 0.85  pF/cm).
Then, for  higher bias voltages,  it  is almost flat,  but the  curves
exhibit a small gradient when the voltage is increased.

According to this observation, the   noise is expected to be   reduced
when increasing the bias voltage.  Any deviation  with respect to this
behaviour   could   indicate  new  phenomena  affecting   the detector
operation, like micro-discharges for  instance, as they  manifest them
self as a sudden onset on the noise.

\subsubsection{Effects of the Bias Voltage and Temperature on the Noise}
\begin{figure}[hbt]
  \centering \mbox{\epsfig{file=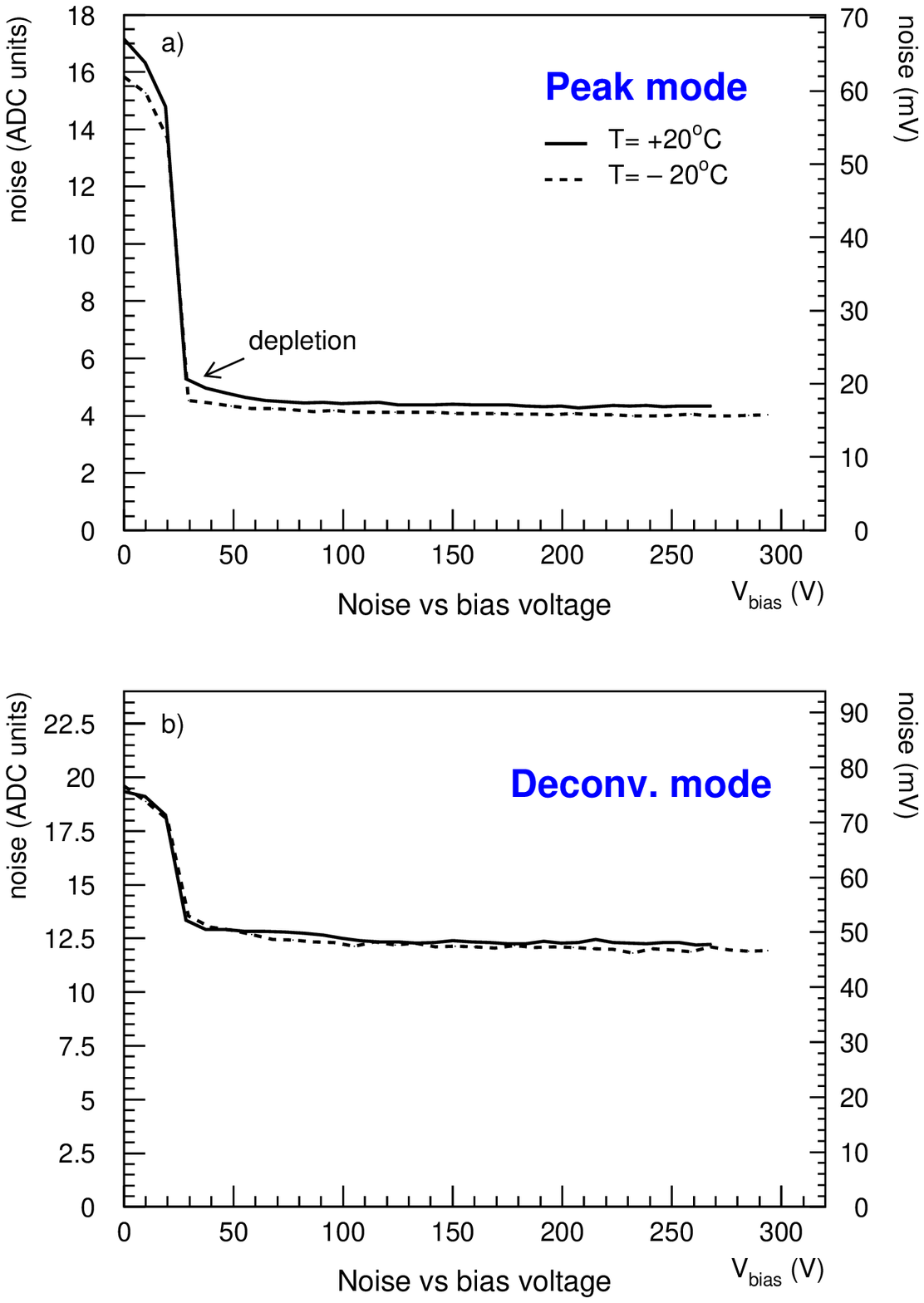,height=21.0cm}}
 \caption{Noise as a function of the detector bias voltage.}\label{fig:sorolla}
\end{figure}

The noise level in the detector has been  studied as a function of the
bias voltage  and the temperature.   The obtained values for both data
sampling modes (peak     and   deconvolution) are shown   in    figure
\ref{fig:sorolla}.  One can   appreciate as the noise  is  drastically
reduced when the  detector is fully   depleted at $\sim 25$V  (in both
data sampling modes).  This behaviour is in perfect agreement with the
observed  one for   the    inter  strip capacitance   (read    section
\ref{sec:is_cap}). The achieved value for the noise is $\sim$ 15 mV in
peak    mode. The Equivalent Noise   Charge  was (ENC)  was $\sim 700$
electrons.

In   figure \ref{fig:sorolla}  a slight   decrease  of the  noise when
increasing the bias voltage can be observed.  The  change is roughly a
10\%  in peak  mode and  a 12\%  in  deconvolution mode.  This can  be
understood as a consequence of the dependence shown by the inter strip
capacitance on the detector bias voltage, and proves that our study is
sensitive to small changes in the detector characteristics (section
\ref{sec:is_cap} and figure \ref{fig:is_cap}).

The noise   of the   detector   exhibits a  small dependence    on the
temperature.  This effect can be observed in figure \ref{fig:sorolla}.
As   expected,  the  noise   level   is  slightly  reduced  at   lower
temperatures.  The  reduction is a  7\% for peak mode  and  a  4\% for
deconvolution    mode.   In  the  other  hand,   studies  of the noise
contribution  from the read out  electronics show that  this is almost
constant with the   temperature.  Consequently the noise  reduction is
due to  the detector contribution, mainly due   to a reduction  of the
leakage current (read section \ref{sec:corrent}).

\subsubsection{No Evidence of Micro-Discharges}
From these  results the production  of micro-discharges in the ATLAS-A
silicon detector prototype is discarded, as any onset  on the noise in
the detector have  been observed, even  at low temperatures  where the
effects are expected to be bigger if micro-discharges occur
\cite{udq-japones}.

\subsection{Hit and Cluster Search}
In  each event  the impact  point   of the  incident  particles on the
detector   is obtained from a strip   cluster  search that proceeds as
follows:
\begin{enumerate}
\item  the strip significance ($s$ =  charge collected in this event /
  noise) is computed for all the channels.
\item Then, the {\sl primary strip} is defined as the channel with the
  highest strip significance if $s>3$.
\item Those strips within an interval of $\pm 5$
  strips around the primary strip with $s>3$ are included in
  the cluster.
\item The  {\sl  hit  significance} is  defined   as the sum    of the
  significance of all strips included in the cluster.
\item  The search for more clusters  in the same event continues until
  no strips with $s>3$ remain.
\end{enumerate}

The  {\sl signal  to noise ratio}  is  taken as  the peak  of the  hit
significance distribution, as it compares  the charge collected in one
strip with its noise and this is done for all the strips in a cluster.

The {\sl cluster size}  is just given by the  number of strips in each
cluster.

\subsection{Signal to Noise Ratio}

The values obtained  in this analysis  for the  signal  to noise ratio
depend basically on the following factors:
\begin{itemize}
\item data sampling (peak or deconvolution mode),
\item bias voltage,
\end{itemize}

Previous studies   using the  FELIX  chip  for analogue  readout tests
\cite{chris_treshold} have shown that  its performance in peak mode is
better than  using  deconvolution mode  (section  \ref{sec:electronics}).
For this reason, only the results obtained in  peak mode are presented
in this section.

For a given bias voltage, temperature and read out scheme the only two
magnitudes  affecting the  signal  to noise  ratio are: the  collected
charge (i.e.  signal) and the noise level.

The collected  charge depends on the incident   particle type.  In the
case  of muons (m.i.p.)  the collected charge  was equivalent to $\sim
19,800   e^-$.  In the   second case,  using the   electrons  from the
radioactive source, the  collected charge was  slightly smaller, $\sim
18,100 e^-$.   The errors associated to  both figures  are $\sim 10\%$
mainly due to the calibration procedure.   The ratio between muons and
the $\sim 2$ MeV electrons  of the  radioactive  source is: $1.09  \pm
0.03$.   In  the  ratio   the   systematic  uncertainties  due  to the
calibration are cancelled.

The collected charge has  been seen not  to depend on the bias voltage
(once the  detector has been  fully depleted).  This can be understood
in terms of a constant efficiency  for collection of the electron-hole
pairs  that    have been created,  once  the   detector has been fully
depleted.  This  reflects that  no     extra charge  is lost in    the
recombination  processes  inside the depleted   zone,  and this is not
expected to be dependent on the bias voltage.

The  noise level has been already  studied in section \ref{sec:noise},
and as explained it is reduced when the bias voltage is increased.

The  signal to  noise    ratio at 30  V   reverse   bias voltage   and
$-20^\circ$C  was   $25.6\pm0.6$  for   muons  and $23.4\pm0.2$    for
electrons.

\begin{figure}[hbt]
  \centering \mbox{\epsfig{file=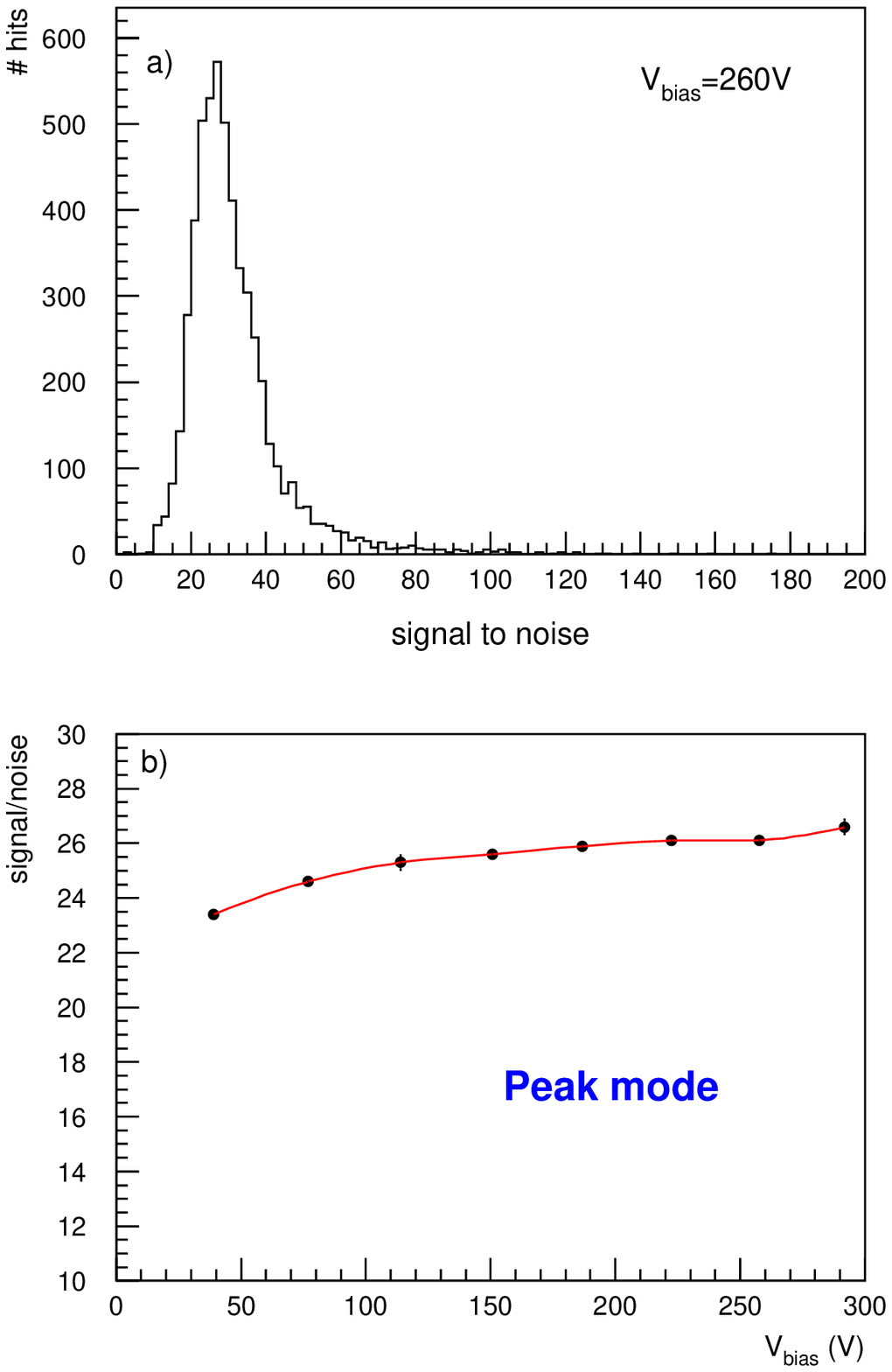,height=21.0cm}}
 \caption{Signal to noise ratio vs. the bias voltage. a) The hit significance 
 distribution when applying a reverse bias voltage of 260 V. b) Evolution of 
 the peak with the reverse bias voltage.}
\label{fig:snr}
\end{figure}

The  amount of charge  collected is the same for  all bias voltage and
the noise  is lower for  higher voltages,  consequently the signal  to
noise  ratio  increases    with  the    bias    voltage  (see   figure
\ref{fig:snr}.b). Using electrons as incident particles, the signal to
noise ratio is $\sim 26:1$.

\subsection{The Cluster Size}

In events with particles crossing   the detector between two   strips,
only a  fraction of  the total produced  charge is  collected  on each
strip.  Thus, the cluster  size is in  an important magnitude  related
with the spatial resolution of  the detector.  The worse resolution is
obtained when  all hits consist in  a  single strip cluster.  Then the
resolution is  just given by:   {\sl  strip pitch}/$\sqrt{12}$.    For
multi-strip clusters, it  could be  improved,  if the  impact point is
just given  by  weighting the  charge collected in   each  strip in  a
suitable way \cite{a.peisert}.

At different  bias  voltages,  the  electric field varies   inside the
detector. This electric  field   accelerates the charges  towards  the
collector strips.   So,  for high values  of  the  electric field, the
expected behaviour is to collect a bigger fraction of  the charge in a
single strip. Then the spatial resolution is degraded.

\begin{figure}[hbt]
  \centering \mbox{\epsfig{file=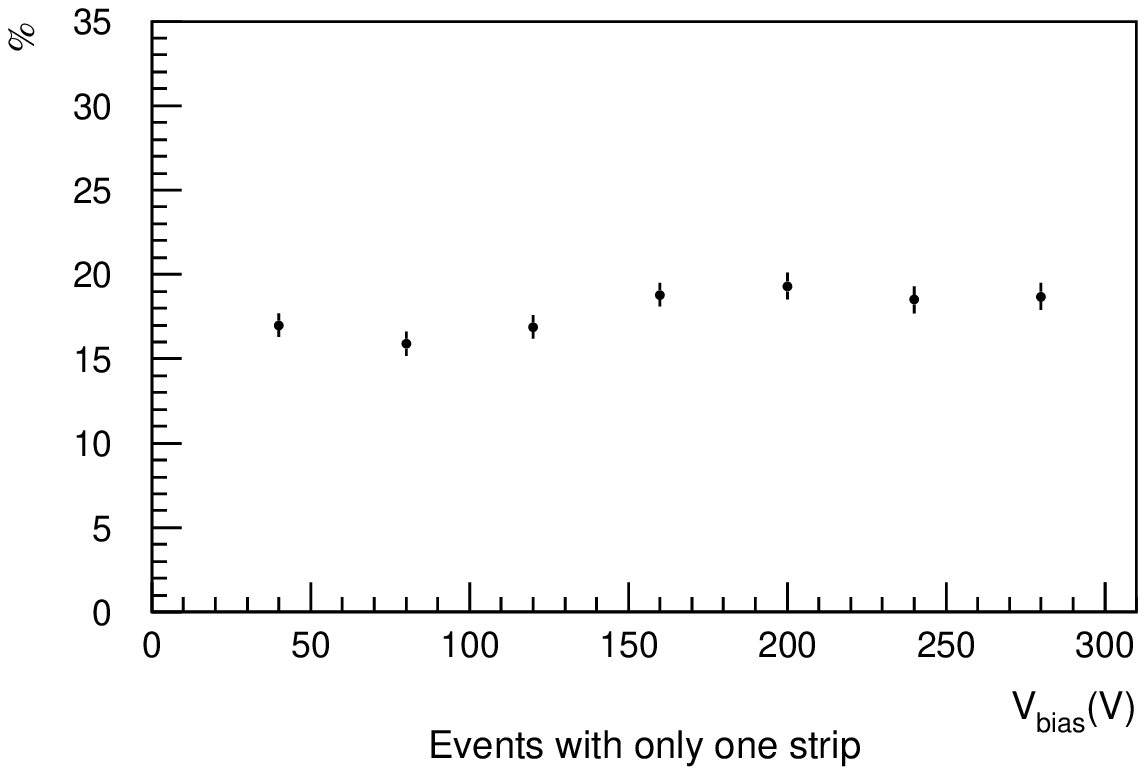,height=10.0cm}}
 \caption{Fraction of events with a single strip cluster as a function of 
 the bias voltage.}
\label{fig:unatira}
\end{figure}

The  results obtained are presented in  figure \ref{fig:unatira}.  The
trend follows  an increase of the  fraction of  clusters with a single
strip when increasing the bias voltage. It  varies from $\sim 16$\% of
the events at low voltage, up to $\sim 19$\% at high biasing voltages.
The absolute change is very  small in a wide   range of voltages.   In
consequence, the spatial resolution is  not dramatically affected when
using high bias voltage values. The fraction of clusters with at least
two strips remains above the 80\%.

The small dependence of the cluster  size with the  bias voltage is in
part due  to the charge   sharing between  the read out   $n^+$-strips
propiciated by the intermediate strip.

\section{Discussion}
With the results presented in the previous section the micro-discharge
production in the ATLAS-A prototype have been discarded.  However, the
leakage current  on that device  presents some peculiarities that need
to be explained.

\subsection{Leakage Current \label{sec:corrent}}
\begin{figure}[htb]
 \centering             \mbox{\epsfig{file=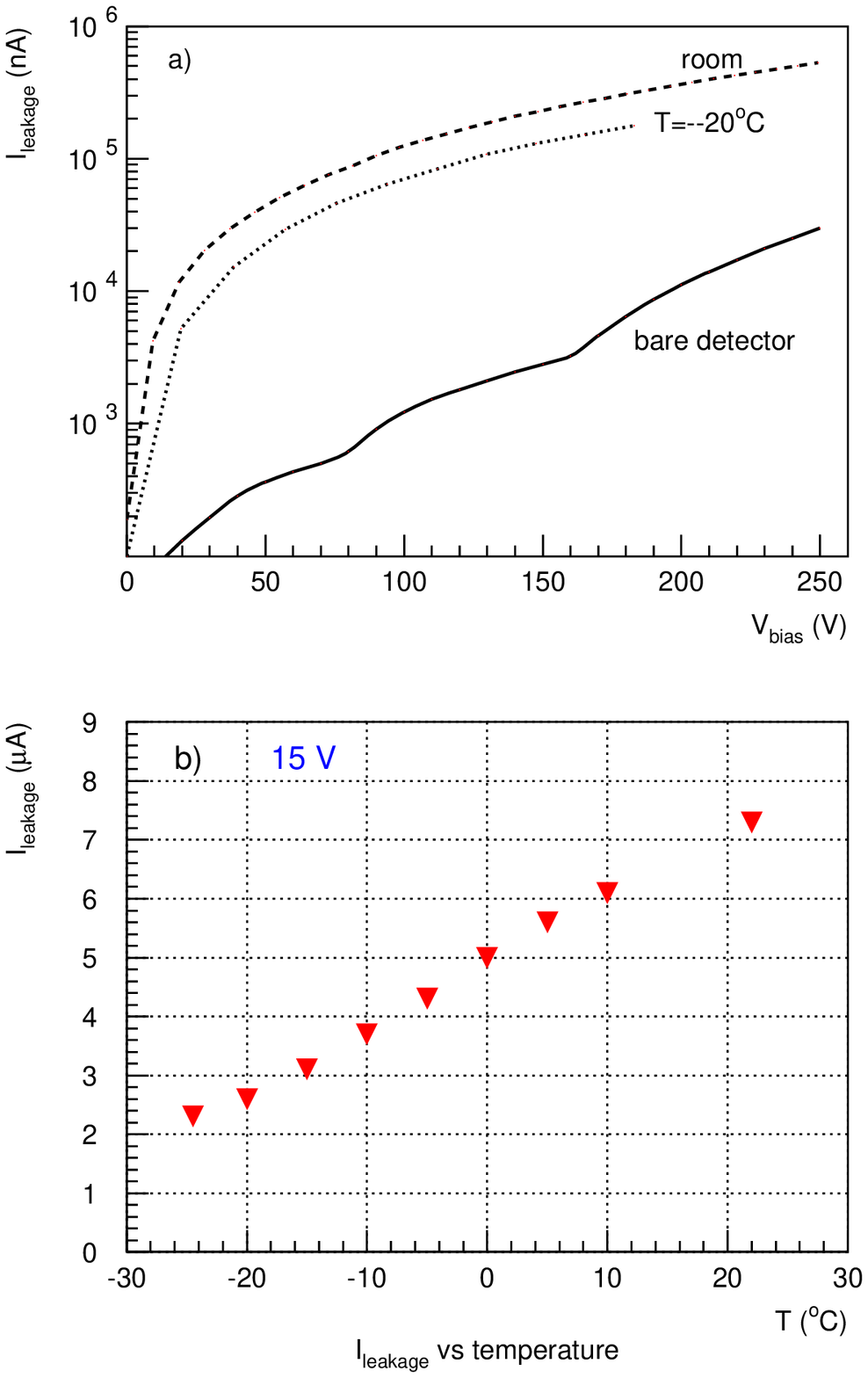,height=19.0cm}}
 \caption{Leakage current in  the detector as a  function of  the bias
 voltage.  The solid line  corresponds to the characterization  of the
 detector  before  bonding.   The   two  others curves   represent the
 measured current  after bonding up  the detector and at two different
 temperatures: room ($\approx 25^o$C) and cooled down ($-20^o$C).  The
 bottom plot shows the current as a  function of the temperature for a
 fix bias voltage.}  \label{fig:corrent}
\end{figure}

The leakage current of the full detector has been studied in each step
of  the assembly process. The measured   values of the leakage current
are shown in  the figure \ref{fig:corrent}.  In  the case  of the bare
detector, two kinks can  be appreciated in  the I-V plot.  These kinks
are   related with a change   in the slope  of the   curve.  These are
reproducible phenomena, and the first   kink (at $\sim 80$V)   matches
quite well with the voltage  for producing micro-discharges  according
to the reference  \cite{udq-japones}.  However, there is no  agreement
in  the   absolute change of   the   current.   Actually in  reference
\cite{udq-japones},  the  micro-discharges  originate a change  bigger
than two orders of magnitude.

Unfortunately, the wire-bonding   process of the  detector channels to
the read out electronics had damaged the prototype due  to the lack of
a passivation layer.  Therefore,  after  wire-bonding, the current  is
increased globally by a factor  $\sim 100$ and  the kink structure  is
lost.

In the other   hand, the  dependence  of the  leakage current  on  the
temperature does not follow the expected one (approx a factor 1/2 each
$8^\circ$C)  as shown in  figure \ref{fig:corrent}.a.  As it is shown,
the current is  reduced just by a factor   $\sim 1/3$ by  changing the
temperature by $\sim   40^\circ$C.  The temperature dependence  of the
leakage current for a fixed bias voltage  is almost linear (see figure
\ref{fig:corrent}.b).

Neither  breakdown  nor    punch-through in the  leakage   current  is
observed, and this is in agreement with a electric field being smaller
than 30 V/$\mu$m, which  is supported by  simulation studies using the
semiconductor simulation package TOSCA \cite{tosca}.

It is believed that: the damage of  the prototype when wire-bonding can
not be responsible of a micro-discharge inhibition in the module.

\section{Conclusions}
The  performance of an  ATLAS-A silicon  micro-strip has been  studied
using a FELIX chip with analogue read out.   The signal to noise ratio
($\sim$26:1) of the detector depends on the reverse bias voltage, as a
consequence of the  noise reduction.  In  the other hand, the  cluster
size shows   that more  than   80\%  of the cluster    are multi-strip
clusters.  This two factors show that a good spatial resolution can be
achieved with the ATLAS-A detector design.

No evidence of   micro-discharges has been   observed in this  device,
neither by burst on the leakage current, nor  by a onset of the noise,
even working at high bias voltages (up to 300 V), and low temperatures
($-25^\circ$C).